\documentclass[aps,prd,nofootinbib,twocolumn,groupedaddress,floatfix]{revtex4}
\usepackage{graphicx}
\usepackage{epsfig}
\usepackage{dcolumn}
\usepackage{amsmath}
\def\Journal#1#2#3#4{{#1} {\bf#2}, #3 (#4)}
\def\NPA{{\rm Nucl. Phys.} A}
\def\NPB{{\rm Nucl. Phys.} B}
\def\PLB{{\rm Phys. Lett.}  B}

\def\PRD{{\rm Phys. Rev.} D}
\def\PRC{{\rm Phys. Rev.} C}

\def\ep{\epsilon}
\def\vep{\varepsilon}
\def\la{\langle}
\def\ra{\rangle}

\def\ka{\kappa}
\def\al{\alpha}

\def\be{\begin{equation}}
\def\ee{\end{equation}}
\def\bea{\begin{eqnarray}}
\def\eea{\end{eqnarray}}
\begin{document}

\title{Electromagnetic Structure of the $\rho$ Meson in the 
Light-Front Quark Model}
\author{ Ho-Meoyng Choi$^a$ and Chueng-Ryong Ji$^b$
\\
$^a$ Department of Physics, Kyungpook National University,
     Daegu, 702-701 Korea\\
$^b$ Department of Physics, North Carolina State University,
Raleigh, NC 27695-8202}

\begin{abstract}
We investigate the elastic form factors of the $\rho$ meson in the
light-front quark model(LFQM). 
With the phenomenologically accessible meson vertices including the one
obtained by the Melosh transformation frequently used in the LFQM,
we find that only the helicity $0\to 0$ matrix element of the plus current 
receives the zero-mode contribution.  We quantify the zero-mode 
contribution in the helicity $0\to 0$ amplitude using the angular 
condition of spin-1 system. After taking care of the zero-mode issue,
we obtain the magnetic($\mu$) and quadrupole($Q$) moments of the $\rho$ 
meson as $\mu=1.92$ and  $Q=0.43$, respectively, 
in the LFQM consistent with the Melosh
transformation and compare our results with other available theoretical 
predictions. 
\end{abstract}
\maketitle
\section{Introduction}
In the past few years, there has been much theoretical 
effort~\cite{BPP,CDKM,Ja99,JC01,BCJ1,BCJ_rho,MS,Ja03,BCJ_PV}
to study the so called ``zero-mode"~\cite{CY,Bur,Melo1,BHw,CJZ,SB}
contribution to the elastic and
weak form factors for spin-0 and 1 systems in light-front dynamics(LFD).
The zero-mode is closely related to the off-diagonal elements in the
Fock state expansion of the current matrix. 
In particular, the zero-mode contribution to the form factor $F(q^2)$ can be 
characterized by the nonvanishing contribution from the off-diagonal elements
in the $q^+\to 0$ limit, where $q^+$ is the longitudinal
component of the momentum transfer $q$( $q^\pm=q^0\pm q^3$,
$q^2=q^+q^--{\bf q}^2_\perp$). 
In the absence of zero-mode, however,
the hadron form factors can be expressed as the convolution
of the initial and final LF wave functions; namely, the physical result
can be obtained by taking into account only the valence contribution or the
diagonal elements in the Fock state expansion.

In this work, we analyze the zero-mode contribution to the elastic form 
factors of the $\rho$ meson using the LF constituent quark model (or LFQM 
in short)~\cite{Ja90,Ja96,Card,Mel96,CJ97,CJ97D,CJMix,CJKaon,KCJ,CJK02,Hwang}, 
which has been quite successful in predicting various electroweak properties 
of light and heavy mesons compared to the available data.  
In our previous LFD analysis~\cite{BCJ_rho} of spin-1 electromanetic form 
factors, we have shown that the zero-mode complication can 
exist even in the matrix element of the plus component of the current.
Using a covariant model of spin-1 system with the polarization vectors
obtained from the LF gauge($\ep^+_{\pm}=0$), we found that only the 
helicity zero-to-zero amplitude, i.e. $(h',h)=(0,0)$, can be contaminated 
by the zero-mode contribution.  However, our conclusion in~\cite{BCJ_rho} 
was based on the use of a rather simple vector
meson vertex $\Gamma^\mu=\gamma^\mu$.  The aim of the present work 
is to explore our findings in the more phenomenologically
accessible $\rho$ meson vertex(see Eq.~(\ref{Gal})).
This includes the case of the $\rho$ meson vertex obtained by the Melosh
transformation frequently used in the LFQM.

Recently, Jaus~\cite{Ja99,Ja03} proposed a covariant light-front approach
that developes a way of including the zero-mode contribution and removing
spurious form factors proportional to the lightlike vector 
$\omega^\mu=(1,0,0,-1)$. Using the LFQM vector meson vertex that we include
in this work, the author in~\cite{Ja03} concluded the existence of a 
zero-mode in the 
form factor $F_2(Q^2)$ of a spin-1 meson. Although his calculation was 
performed in a covariant way not using the helicity components, the form factor
$F_2$ is related with the matrix element of $(h',h)=(+,0)$-component of 
the current(see Eq.~(2.6) in~\cite{Ja03}) as well as $(h',h)=(0,0)$ component.
Thus, his result indicates that both helicity $(+,0)$ and $(0,0)$ amplitudes
receive the zero-mode contribution. In the present work, 
we do not 
agree~\footnote{We encountered previously a similar situation of 
disagreement in the zero-mode contribution to the $A_1$(or $f$) form factor
for the transition between pseudoscalar and vector mesons.
However, after the compleletion of the work~\cite{BCJ_PV}, the author of
Ref.~\cite{Ja03} informed us in a private communication that he completely
agrees with our results~\cite{BCJ_PV} due to the identity Eq.~(3.33) of
Ref.~\cite{Ja03}. Nonetheless, the communication on the spin-1 elastic form
factors presented in this work has not yet been made.}
with this result but find the zero-mode contribution only in the helicity
$(h',h)=(0,0)$ amplitude.  This result is quite significant in the LFQM
phenomenology because the absence of the zero-mode contamination in the
helicity $(+,0)$ amplitude can give a tremendous benefit in making 
reliable
predictions on the spin-1 observables as we present in the example of the
$\rho$ meson. Our work should be intrinsically distinguished from the 
formulation involving $\omega$, since our formulation involves neither
$\omega$ nor any unphysical form factor.

The paper is organized as follows. In Sec. II, we present the 
Lorentz-invariant electromagnetic form factors and the kinematics for
the reference frames used in our analysis.
We also discuss the LF helicity basis, the angular condition for spin-1
systems, and the two particular prescriptions
used in extracting the physical form factors.
In Sec. III, we present our LF calculation varying the vector meson vertex. 
We employ both the manifestly covariant model vertex(beyond the simple
model of $\Gamma^\mu=\gamma^\mu$) and the LFQM vertex consistent with the 
Melosh transformation.  
We discuss the LF valence and nonvalence contributions
using a plus component of the current and show that only the 
helicity zero-to-zero amplitude receives zero-mode contribution 
for the employed meson vertices.  
In Sec. IV, we present our numerical results for the physical quantities
of the $\rho$ meson using the LFQM and compare our results with other 
available theoretical predictions. 
Conclusions follow in Sec. V. In Appendix A, we summarize our
results of the trace calculations for various helicity components used in
our form factor calculations.
\section{Electromagnetic  structure of spin-1 system}

The Lorentz-invariant electromagnetic form factors $F_i(i=1,2,3)$
for a spin-1 particle are defined~\cite{ACG} 
by the matrix elements of the current $J^\mu$ between the initial 
$|P,h\ra$(momentum $P$ and helicity $h$) and the final
$|P',h'\ra$ eigenstates as follows:
\bea\label{F123}
\hspace{-0.5cm}\la P',h'|J^\mu|P,h\ra
&=& -\ep^{*}_{h'}\cdot\ep_h(P+P')^\mu F_1(Q^2)\nonumber\\
&+& (\ep^\mu_h\;q\cdot\ep^{*}_{h'} 
- \ep^{*\mu}_{h'}\;q\cdot\ep_h)F_2(Q^2)\nonumber\\
&+& \frac{(\ep^{*}_{h'}\cdot q)(\ep_h\cdot q)}{2M^2_v} 
(P+P')^\mu F_3(Q^2),
\eea
where $q=P'-P$ and $\ep_h[\ep_{h'}]$ is 
the polarization vector of the initial[final] meson with the physical
mass $M_v$.

We analyze the matrix elements in the Breit 
frame( $q^+=0, q_y=0, q_x=Q$, and ${\bf P}_\perp=-{\bf P'}_\perp$)
defined by~\cite{GK,CCKP,BH,Kei,Card,CJ97}
\bea\label{BF_F}
q^\mu&=&(0,0,Q, 0)\nonumber\\
P^\mu&=&(M_v\sqrt{1+\ka}, M_v\sqrt{1+\ka}, -Q/2,0)\nonumber\\
P'^\mu&=&(M_v\sqrt{1+\ka}, M_v\sqrt{1+\ka}, Q/2,0),
\eea
where $\kappa=Q^2/4M_v^2$ is the kinematic factor. Here, we use the 
notation $p=(p^+,p^-,p^1,p^2)$ and the metric
convention $p^2=p^+p^- -{\bf p}^2_\perp$ with $p^\pm=p^0\pm p^3$.

Following Bjorken-Drell convention~\cite{BD},
we work with the circular polarization and 
spin projection $h=\pm=\uparrow\downarrow$.
With $\ep^+(p,\pm)=0$ from the LF gauge~\cite{BPP}, the polarization
vector is given by

\be\label{BD}
\ep^\mu(p,\pm)=\biggl(0,
\frac{2{\vec\ep}_\perp(\pm)\cdot{\vec p}_\perp}{p^+},
{\vec\ep}_\perp(\pm)\biggr),
\ee
which satisfies $p\cdot\ep(\pm)=0$ with
${\vec\ep}_\perp(\pm)=\mp 1/\sqrt{2}(1,\pm i)$.
Effectively, the initial and final transverse($h=\pm$) and
longitudinal($h=0$) polarization vectors in the
Breit frame are given by
\bea\label{pol_tl}
\ep^\mu(P,\pm)&=&
\frac{\mp 1}{\sqrt{2}}\biggl(0, \frac{-Q}{P^+}, 1, \pm i\biggr),
\nonumber\\
\ep^\mu(P,0)&=&
\frac{1}{M_v}\biggl(P^+, \frac{-M^2_v+ Q^2/4}{P^+},
\frac{-Q}{2}, 0\biggr),
\nonumber\\
\ep^\mu(P',\pm)&=&
\frac{\mp 1}{\sqrt{2}}\biggl(0, \frac{Q}{P^+}, 1, \pm i\biggr),
\nonumber\\
\ep^\mu(P',0)&=&
\frac{1}{M_v}\biggl(P^+, \frac{-M^2_v+ Q^2/4}{P^+},
\frac{Q}{2}, 0\biggr).
\eea
The covariant form factors of a spin-1 hadron in Eq.~(\ref{F123})
can be determined using only the plus component of the current,
$I^+_{h'h}(0)\equiv\la P',h'|J^\mu|P,h\ra$.
As one can see from Eq.~(\ref{F123}), while there are only three 
independent(invariant)
form factors, nine elements of $I^+_{h'h}(0)$ can be assigned
to the current operator.
However, since the current matrix elements  $I^+_{h'h}(0)$ must be
constrained by the invariance under 
the LF parity and the LF time-reversal,
one can reduce the independent matrix elements of the current
down to four, e.g. $I^+_{++},I^+_{+-},I^+_{+0}$ and 
$I^+_{00}$~\cite{GK,CCKP,BH,Card,CJ97}. The angular momentum conservation 
requires an additional constraint on the current operator, which yields
the so called ``angular condition" $\Delta(Q^2)$ given by~\cite{GK}
\be\label{ac}
\Delta(Q^2)=(1+2\kappa) I^+_{++} + I^+_{+-}
-\sqrt{8\kappa}I^+_{+0} - I^+_{00}=0.
\ee
Because of the angular condition, 
only three helicity amplitudes are independent as expected from the
three physical form factors.
However, the relations between the physical form factors
and the matrix elements $I^+_{h'h}$ are not uniquely determined
because the number of helicity amplitudes is larger than that of physical 
form factors. 
So one may choose which matrix elements to use to extract the form
factors.  For example, Grach and Kondratyuk(GK)~\cite{GK}
used only $(h',h)=(+,+),(+,-)$ and ($+,0$), but not the pure
longitudinal (0,0) component. On the other hand, Brodsky and 
Hiller(BH)~\cite{BH} used $(0,0),(+,0)$, and ($+,-$) amplitudes 
considering that the (0,0) component gives the most dominant
contribution in the high momentum perturbative QCD(pQCD) region.
Chung, Coester, Keister and Polyzou(CCKP)~\cite{CCKP} involved all 
helicity states, i.e. $(+,+),(0,0),(+,0)$, and $(+,-)$.
Among these various choices, we present GK and BH prescriptions for the
comparison purpose in this work.  In practical 
computation, instead of calculating the Lorentz-invariant form
factors $F_i(Q^2)$, the physical charge($G_C$), magnetic($G_M$),
and quadrupole($G_Q$) form factors are often used.
The relation between $F's$ and $G's$ are given by
\bea\label{FG}
G_C&=& F_1 + \frac{2}{3}\kappa G_Q,\nonumber\\
G_M&=& - F_2, \nonumber\\
G_Q&=& F_1 + F_2 + (1+\kappa)F_3.
\eea
The physical form factors($G_C,G_M$, and $G_Q$) in terms of the matrix 
elements $I^+_{h'h}$ for GK and BH prescriptions are given by
%\footnote{The form factors ($G_0,G_1,G_2$) used in Ref.~\cite{Card} and 
%the ones used in this work are related by $G_0=2P^+G_C$, $G_1=2P^+G_M$, and
%$G_2=(\sqrt{8}\kappa/3)2P^+G_Q$.} 
\bea\label{G_GK}
G^{\rm GK}_{C}&=&\frac{1}{2P^+}\biggl[
\frac{(3-2\kappa)}{3}I^+_{++} 
+\frac{4\kappa}{3}\frac{I^+_{+0}}{\sqrt{2\kappa}}
+ \frac{1}{3}I^+_{+-}\biggr],
\nonumber\\
G^{\rm GK}_M&=&\frac{2}{2P^+}\biggl[I^+_{++}
-\frac{1}{\sqrt{2\kappa}}I^+_{+0}\biggr],
\nonumber\\
G^{\rm GK}_Q&=&\frac{1}{2P^+}\biggl[ -I^+_{++}
+ 2\frac{I^+_{+0}}{\sqrt{2\kappa}}-\frac{I^+_{+-}}{\kappa}\biggr],
\eea
and 
\bea\label{G_BH}
G^{\rm BH}_C &=& \frac{1}{2P^+(1+2\kappa)}
\biggl[ \frac{3-2\kappa}{3}I^+_{00} 
+ \frac{16}{3}\kappa\frac{I^+_{+0}}{\sqrt{2\kappa}}
\nonumber\\
&&\hspace{2.5cm}+ \frac{2}{3}(2\kappa-1)I^+_{+-} \biggr],
\nonumber\\
G^{\rm BH}_M &=& \frac{2}{2P^+(1+2\kappa)}
\biggl[ I^+_{00} + \frac{(2\kappa -1)}{\sqrt{2\kappa}}I^+_{+0} 
- I^+_{+-}\biggr],
\nonumber\\
G^{\rm BH}_Q &=& \frac{-1}{2P^+(1+2\kappa)}
\biggl[I^+_{00} - 2\frac{I^+_{+0}}{\sqrt{2\kappa}} 
+ \frac{1+\kappa}{\kappa}I^+_{+-} \biggr].
\eea
Of course, the matrix elements must fulfull the angular condition 
given by Eq.~(\ref{ac}) for the physical form factors to be independent from
the prescriptions(GK or BH).

At zero momentum transfer, these form factors are proportional to the
usual static quantities of charge $e$, magnetic moment $\mu$,
and quadrupole moment $Q$~\cite{Kei,Card}:
\be\label{Q_1}
eG_C(0) = e, \;\; eG_M(0)=2M_v\mu, \;\; eG_Q(0)=M^2_vQ.
\ee
In principle, these physical form factors are also related to the structure
functions $A(Q^2)$, $B(Q^2)$ and the tensor polarization $T_{20}$~\cite{BH}:
\begin{widetext}
\bea\label{AB}
A(Q^2)&=& G^2_C + \frac{2}{3}\kappa G^2_M 
+ \frac{8}{9}\kappa^2G^2_Q,\;\;
B(Q^2)=\frac{4}{3}\kappa(1+\kappa)G^2_M,
\nonumber\\
T_{20}(Q^2,\theta)
&=&-\kappa\frac{\sqrt{2}}{3}
\frac{\frac{4}{3}\kappa G^2_Q + 4G_QG_C
+[1/2 + (1+\kappa)\tan^2(\theta/2)]G^2_M}
{A + B\tan^2(\theta/2)}.
\eea
\end{widetext}
\section{Model description and calculation}
As we have shown in our previous work~\cite{BCJ_rho},
the form factors of a spin-1 particle can be derived
from the covariant Bethe-Salpeter(BS) model of 
($3+1$)-dimensional fermion field theory.

\begin{figure*}
\includegraphics[width=6in]{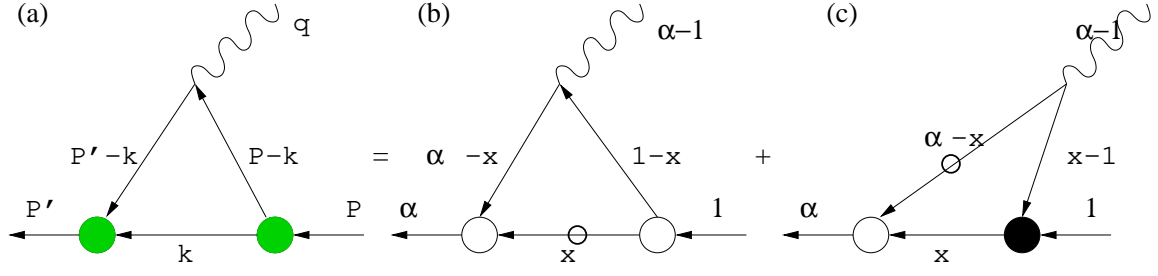}
\caption{Covariant triangle diagram (a) is represented as the sum
of LF valence diagram (b) defined in $0<k^+<P^+$ region and the 
nonvalence diagram (c) defined in $P^+<k^+<P'^+$ region, where
$\al=P'^+/P^+=1+q^+/P^+$. The large white and black blobs at
the meson-quark vertices in (b) and (c) represent the ordinary
LF wave function and the nonvalence wave function vertices, respectively.
The small white blobs in (b) and (c) represent the (on-shell)
mass pole of the quark propagator determined from the $k^-$-integration. 
\label{Fig1}}
\end{figure*}

The covariant diagram shown in Fig.~\ref{Fig1}(a) is in general
equivalent to the sum of LF valence diagram (b) and the 
nonvalence diagram (c), where $\al=P'^+/P^+=1+q^+/P^+$.
The electromagnetic(EM) current $I^\mu_{h'h}(0)$ of
a spin-1 particle with equal constituent($m=m_q=m_{\bar q}$)
obtained from the covariant diagram of  
Fig.~\ref{Fig1}(a) is given by
\begin{widetext}
\be\label{jji}
I^\mu_{h'h}(0)=
iN_cg^2\int\frac{d^4k}{(2\pi)^4}
\frac{S_\Lambda(k-P)S^\mu_{h'h}S_\Lambda(k-P')}
{[(k-P)^2-m^2+i\vep][k^2-m^2+i\vep][(k-P')^2-m^2+i\vep]},
\ee
\end{widetext}
where $N_c$ is the number of colors
and $g$ is the normalization constant modulo the charge factor $e_q$
which can be fixed by requiring the charge form factor to be unity at
$q^2=0$.  In Ref.~\cite{BCJ_rho},
we replace the point photon-vertex $\gamma^\mu$ by a nonlocal smeared
photon-vertex $S_\Lambda(P)\gamma^\mu S_\Lambda(P')$
to regularize the covariant fermion triangle loop in ($3+1$) dimensions, where
$S_\Lambda(P)=\Lambda^2/(P^2-\Lambda^2+i\varepsilon)$ and $\Lambda$ plays
the role of a momentum cutoff similar to the Pauli-Villars
regularization.

The trace term $S^\mu_{h'h}$ of the 
quark propagators in Eq.~(\ref{jji}) is given by
\be\label{Sji}
S^\mu_{h'h}=(\ep^{*}_{h'})_\al{\rm Tr}
[\Gamma^\al(\not\!p'+m)\gamma^\mu
(\not\!p+m)\Gamma^\beta(\not\!k + m)](\ep_h)_\beta,
\ee
where $p=k-P,p'=k-P'$ and $\Gamma^\beta(\Gamma^\al)$ is the spinor 
structure of initial(final) state meson-quark vertex. 
In our previous work~\cite{BCJ_rho}, 
using a rather simple meson vertex, $\Gamma^\mu=\gamma^\mu$, 
we showed that only $S^+_{00}$ receives zero-mode contributions.

The purpose of the present work is to go beyond the simple vertex,
$\Gamma^\mu=\gamma^\mu$, and analyze the zero-mode contribution
to see if the same conclusion(i.e. zero-mode only in $S^+_{00}$)
can be drawn.
For this purpose, we extend the meson vertex to the more general one,
which has been used for the phenomenology: 
\be\label{Gal}
\Gamma^\beta=\gamma^\beta - \frac{(2k-P)^\beta}{D},\;\;
\Gamma^\al=\gamma^\al - \frac{(2k-P')^\al}{D'}.
\ee
While in the manifestly covariant model, $D$ and $D'$ may be 
written as~\cite{MT,MF03}~\footnote{In a brief presentation of 
Ref.~\cite{MF03},
the authors indicated that they investigated the rho meson form
factors using Eq.~(\ref{aap_cov}). However, the detailed explicit
calculations have not yet been presented. Moreover, the result shown
in Ref.~\cite{MF03} was based on the instant-form polarization rather
than the light-front polarization so that the direct analysis of 
light-front helicity amplitudes as discussed in this work has not
yet been presented.}
\bea\label{aap_cov}
D_{\rm cov}&=& \frac{2}{M_v}(k\cdot P+M_vm-i\ep),\nonumber\\
D'_{\rm cov}&=& \frac{2}{M_v}(k\cdot P'+M_vm-i\ep),
\eea
in the standard LFQM they are given by~\cite{Card,Ja99,Ja03}
\bea\label{aap_LF}
D_{\rm LFQM}&=& \sqrt{\frac{m^2+{\bf k}^2_{i\perp}}{x(1-x)} } + 2m
         = M_{0i} + 2m,\nonumber\\
D'_{\rm LFQM}&=& \sqrt{\frac{m^2+{\bf k}^2_{f\perp}}{x'(1-x')} } + 2m
         = M_{0f} + 2m,
\eea
where $x'=x/\al$, $\al=P'^+/P^+=1+q^+/P^+$ (i.e. $x'=x$ in $q^+=0$ frame) 
and 
\be\label{kperp}
{\bf k}_{i\perp}={\bf k}_\perp + \frac{x}{2}{\bf q}_\perp,\;
{\bf k}_{f\perp}={\bf k}_\perp - \frac{x'}{2}{\bf q}_\perp.
\ee
We note that the LF meson vertex(Eq.~(\ref{Gal})) with
$D_{\rm LFQM}$ given by Eq.~(\ref{aap_LF}) can be
obtained by the standard LF Melosh transformation~\cite{Ja90}.
The equivalence between $D_{\rm cov}(D'_{\rm cov})$ 
in Eq.~(\ref{aap_cov}) and $D_{\rm LFQM}(D'_{\rm LFQM})$ in Eq.~(\ref{aap_LF}) 
can be established only in a very limited case, e.g. 
zero binding energy limit, $M_v=M_0$. 

As we have shown in our previous analysis, with a rather simple but
manifestly covariant meson 
vertex($\Gamma^\mu=\gamma^\mu$)~\cite{BCJ_rho,BCJ_PV}, we can in principle
duplicate both the manifestly covariant calculation and the LF calculation to 
check which form factors are immune to the zero-mode. However, as we 
have also shown in Refs.~\cite{BCJ_rho,BCJ_PV}, we can directly check
the power counting of the longitudinal momentum fraction in the trace
term to find out explicitly which helicity amplitude has the zero-mode
contribution. No matter which way we check, the results are the same.
In this work, we follow the latter procedure to check each helicity
amplitude and show 
explicitly that both phenomenologically accessible meson vertices given
by Eqs.~(\ref{aap_cov}) and~(\ref{aap_LF}) lead to the same conclusion,
i.e. the zero-mode contribution exists only in the helicity zero-to-zero
amplitude.

Before we proceed, we first summarize our result of the trace term for
the `$+$'-component of the current. Then, we separate the valence contribution
from the nonvalence contribution and take the $q^+\to 0$ limit of the 
nonvalence contribution to analyze the zero-mode in each helicity 
amplitude.

\subsection{ Trace Calculation} 
From Eqs.~(\ref{Sji}) and~(\ref{Gal}), we obtain the trace term
for the $'+'$-component of the current as follows
%\begin{widetext}
\bea\label{Sji_4}
&&\hspace{-0.5cm}S^+_{h'h}
\nonumber\\
&=&(\ep^{*}_{h'})_\al{\rm Tr}
[\Gamma^\al(\not\!p'+m)\gamma^+
(\not\!p+m)\Gamma^\beta(\not\!k + m)](\ep_h)_\beta,\nonumber\\
&=& {\rm Tr}
[\not\!\ep^{*}_{h'}(\not\!p'+m)\gamma^+
(\not\!p+m)\not\!\ep_h(\not\!k + m)]\nonumber\\
&&- \frac{\ep_h\cdot(2k-P)}{D}
{\rm Tr}[\not\!\ep^{*}_{h'}(\not\!p'+ m)\gamma^+
(\not\!p+m)(\not\!k + m)],\nonumber\\
&&- \frac{\ep^*_{h'}\cdot(2k-P')}{D'}
{\rm Tr}[(\not\!p'+ m)\gamma^+
(\not\!p+m)\not\!\ep_h(\not\!k + m)],\nonumber\\
&&+ \frac{\ep^*_{h'}\cdot(2k-P')}{D'}
\frac{\ep_h\cdot(2k-P)}{D}
\nonumber\\
&&\;\;\times
{\rm Tr}[(\not\!p'+ m)\gamma^+ (\not\!p+m)(\not\!k + m)],\nonumber\\
&=& S^+_{1h'h} - S^+_{2h'h} - S^+_{3h'h}
+ S^+_{4h'h}.
\eea
%\end{widetext}
Using the following identity
\be\label{on_inst}
\not\!p + m = (\not\!{p_{\rm on}} + m)
+ \frac{1}{2}\gamma^+(p^- - p^-_{\rm on}),
\ee
we separate the trace part for each $S^+_{ih'h}(i=1,2,3,4)$ into the 
on-mass shell(i.e. $p^2=m^2$ or $p^-=p^-_{\rm on}=(m^2+{\bf p}^2_\perp)/p^+$)
propagating part, $(S^+_{ih'h})_{\rm on}$,
and the off-shell part, $(S^+_{ih'h})_{\rm off}$,
i.e.  $S^+_{ih'h}=(S^+_{ih'h})_{\rm on} +(S^+_{ih'h})_{\rm off}$. 

The on-mass shell parts $(S^+_{ih'h})_{\rm on}$ are
given by
\bea\label{Sji_on}
(S^+_{1h'h})_{\rm on}&=&{\rm Tr}
[\not\!\ep^{*}_{h'}(\not\!p'_{\rm on}+m)\gamma^+
(\not\!p_{\rm on}+m)\not\!\ep_h(\not\!k_{\rm on} + m)],
\nonumber\\
(S^+_{2h'h})_{\rm on}&=&\frac{\ep_h\cdot(2k_{\rm on}-P)}{D}
\nonumber\\
&&\times
{\rm Tr}[\not\!\ep^{*}_{h'}(\not\!p'_{\rm on}+ m)\gamma^+
(\not\!p_{\rm on}+m)(\not\!k_{\rm on} + m)],
\nonumber\\
(S^+_{3h'h})_{\rm on}&=&\frac{\ep^*_{h'}\cdot(2k_{\rm on}-P')}{D'}
\nonumber\\
&&\times
{\rm Tr}[(\not\!p'_{\rm on}+ m)\gamma^+
(\not\!p_{\rm on}+m)\not\!\ep_h(\not\!k_{\rm on} + m)],
\nonumber\\
(S^+_{4h'h})_{\rm on}&=&\frac{\ep^*_{h'}\cdot(2k_{\rm on}-P')}{D'}
\frac{\ep_h\cdot(2k_{\rm on}-P)}{D}
\nonumber\\
&&\times
{\rm Tr}[(\not\!p'_{\rm on}+ m)\gamma^+
(\not\!p_{\rm on}+m)(\not\!k_{\rm on} + m)],
\eea
and the off-shell parts $(S^+_{ih'h})_{\rm off}$ are
given by
\bea\label{Sji_inst}
(S^+_{1h'h})_{\rm off} &=&
\frac{(k^- - k^-_{\rm on})}{2}
\nonumber\\
&&\times
{\rm Tr}[\not\!\ep^{*}_{h'}(\not\!{p'_{\rm on}} + m)\gamma^+
(\not\!{p_{\rm on}}+m)\not\!\ep_h\gamma^+],
\nonumber\\
(S^+_{2h'h})_{\rm off} &=&
\frac{\ep^+_h(k^--k_{\rm on}^-)}{D}
\nonumber\\
&&\times
{\rm Tr}[\not\!\ep^{*}_{h'}(\not\!p'_{\rm on}+ m)\gamma^+
(\not\!p_{\rm on}+m)(\not\!k_{\rm on} + m)]
\nonumber\\
&&+\frac{(k^- - k^-_{\rm on})}{2D}
\ep_h\cdot(2k-P)
\nonumber\\
&&\times
{\rm Tr}[\not\!\ep^{*}_{h'}(\not\!{p'_{\rm on}} + m)\gamma^+
(\not\!{p_{\rm on}}+m)\gamma^+],
\nonumber\\
(S^+_{3h'h})_{\rm off} &=&
\frac{\ep^{*+}_{h'}(k^--k_{\rm on}^-)}{D'}
\nonumber\\
&&\times
{\rm Tr}[(\not\!p'_{\rm on}+ m)\gamma^+
(\not\!p_{\rm on}+m)\not\!\ep_h(\not\!k_{\rm on} + m)]
\nonumber\\
&&+\frac{(k^- - k^-_{\rm on})}{2D'}
\ep^*_{h'}\cdot(2k-P')
\nonumber\\
&&\times
{\rm Tr}[(\not\!{p'_{\rm on}} + m)\gamma^+
(\not\!{p_{\rm on}}+m) \not\!\ep_{h}
\gamma^+],
\nonumber\\
(S^+_{4h'h})_{\rm off} &=&
\frac{(k^--k_{\rm on}^-)}{DD'}
[\epsilon^+_h\epsilon^*_{h'}\cdot(2k_{\rm on}-P')
\nonumber\\
&&+ \epsilon^{*+}_{h'}\epsilon_h\cdot(2k_{\rm on}-P)
+\epsilon^+_h\epsilon^{*+}_{h'}(k^--k^-_{\rm on})]
\nonumber\\
&&\times
{\rm Tr}[(\not\!p'_{\rm on}+ m)\gamma^+
(\not\!p_{\rm on}+m)(\not\!k_{\rm on} + m)]
\nonumber\\
&&+\frac{(k^- - k^-_{\rm on})}{2DD'}
\ep^*_{h'}\cdot(2k-P')\ep_h\cdot(2k-P)
\nonumber\\
&&\times
{\rm Tr}[(\not\!{p'_{\rm on}} + m)\gamma^+
(\not\!{p_{\rm on}}+m)\gamma^+].
\eea
In Appendix A, we present the explicit expressions of
($S^+_{ih'h}$) in terms of LF variables.

Now, by doing the integration over $k^-$ in Eq.~(\ref{jji}), one finds
the two LF time-ordered contributions to the residues 
corresponding to the two poles in $k^-$, the one coming from the interval
(I) $0<k^+<P^+$[see Fig.~\ref{Fig1}(b), the ``valence contribution"],
and the other from (II) $P^+<k^+<P'^+$[see Fig.~\ref{Fig1}(c),
the ``nonvalence contribution"]. 

\subsection{Valence contribution}
In the valence region of $0<k^+<P^+$ as shown in
Fig.~\ref{Fig1}(b), the residue is at the pole of
$k^-=k^-_{\rm on}=[m^2+{\bf k}^2_\perp -i\vep]/k^+$(i.e. spectator
quark), which is placed in the lower half of complex-$k^-$ plane.
Thus, the Cauchy integration over $k^-$ of the plus current,
$I^+_{h'h}(0)$(See Eq.~(\ref{jji})), in the Breit frame 
given by Eq.~(\ref{BF_F}) yields 
%\begin{widetext}
\bea\label{jjiv}
I^{+{\rm val}}_{h'h}&=& \frac{N_c}{16\pi^3}
\int^{1}_{0}\frac{dx}{x(1-x)^2}
\nonumber\\
&\times&\int d^2{\bf k}_\perp
\chi_i(x,{\bf k}_{i\perp}) S^+_{h'h}
\chi_f(x,{\bf k}_{f\perp}),
\eea
%\end{widetext}
where
\bea\label{chii}
\chi_i(x,{\bf k}_{i\perp})&=&
\frac{ g\Lambda^2}{(1-x)(M^2_v-M^2_{0i})(M^2_v-M^2_{0\Lambda i})}
\eea
corresponds to the initial state meson vertex function with 
%${\bf k}_{i\perp}={\bf k}_\perp - x{\bf P}_\perp$  and 
\be\label{inmass}
M^2_{0\Lambda i}=\frac{m^2+ {\bf k}^2_{i\perp}}{x}
       + \frac{\Lambda^2+ {\bf k}^2_{i\perp}}{1-x}.
\ee
The final state
denoted by subscript $(f)$ can be obtained by replacing 
$(x,{\bf k}_{i\perp})$
with $(x',{\bf k}_{f\perp})$ in Eqs.~(\ref{chii}) and~(\ref{inmass}).
Since $k^-=k^-_{\rm on}$ in the valence region,
$(S^+_{h'h})_{\rm off}=0$ and $S^+_{h'h}=(S^+_{h'h})_{\rm on}$.

\subsection{Nonvalence contribution and zero-mode in $q^+\to 0$ limit}

In the region $P^+<k^+<P'^+(=P^++q^+)$ as shown in Fig.~\ref{Fig1}(c),
the poles are at 
$k^- \equiv k^-_m= P'^-_1+[m^2+({\bf k}_\perp -{\bf P'}_\perp)^2
-i\varepsilon]/(k^+ - P'^+)$
(from the struck quark propagator)
and $k^-\equiv k^-_\Lambda = 
P'^-+[\Lambda^2+({\bf k}^2_\perp-{\bf P'}_\perp)^2-i\varepsilon]
/(k^+-P'^+)$( from the smeared quark-photon vertex $S_\Lambda(k-P')$).
Both of them are located in the upper half of the complex $k^-$ plane.

When we do the Cauchy integration over $k^-$ to obtain the LF time-ordered
diagrams, we decompose the product of five energy denominators in
Eq.~(\ref{jji}) into a sum of terms with three energy denominators(see
Eq.~(21) in Ref.\cite{BCJ_rho}) to avoid the complexity of treating double
$k^-$ poles and obtain
\begin{widetext}
\bea\label{jnon}
I^{+nv}_{h'h}&=&-\frac{g^2\Lambda^4}{16\pi^3(\Lambda^2-m^2)^2}
\int^\al_1 \frac{dx}{xx''(x-\al)}\int d^2{\bf k}_\perp
\biggl\{
\frac{S^+_{h'h}(k^-=k^-_\Lambda)}
{(M^2_v-M^2_{0\Lambda f})(q^2-M^2_{\Lambda\Lambda})}
-\frac{S^+_{h'h}(k^-=k^-_\Lambda)}
{(M^2_v-M^2_{0\Lambda f})(q^2-M^2_{\Lambda m})}
\nonumber\\
&&+
\frac{S^+_{h'h}(k^-=k^-_m)}
{(M^2_v - M^2_{0f})(q^2-M^2_{mm})}
-\frac{S^+_{h'h}(k^-=k^-_m)}
{(M^2_v - M^2_{0f})(q^2-M^2_{m\Lambda})}
\biggr\},
\eea
\end{widetext}
where 
\bea\label{inmass_nv}
M^2_{\Lambda\Lambda}&=&
\frac{{\bf k''}^2_\perp+\Lambda^2}{x''(1-x'')},
\;
M^2_{mm}=
\frac{{\bf k''}^2_\perp+m^2}{x''(1-x'')},
\nonumber\\
M^2_{\Lambda m}&=&
\frac{{\bf k''}^2_\perp+\Lambda^2}{x''}
+ \frac{{\bf k''}^2_\perp+m^2}{1-x''},
\nonumber\\
M^2_{m\Lambda}&=&
\frac{{\bf k''}^2_\perp+m^2}{x''}
+ \frac{{\bf k''}^2_\perp+\Lambda^2}{1-x''},
\eea
with the variables defined by
\bea\label{xpp}
x''&=& \frac{1-x}{1-\al},\;
{\bf k''}_\perp={\bf k}_\perp + \biggl(\frac{1}{2}-x''\biggr){\bf q}_\perp.
\eea
Now, the zero-mode~\cite{BCJ_rho}
appears if the nonvalence diagram does not vanish in $q^+\to 0$ limit, i.e.
\bea\label{limit}
\lim_{q^+\to 0}\int^{P^++q^+}_{P^+}dk^+(\cdots)
\equiv\lim_{\al\to 1}\int^{\al}_1 dx(\cdots)\neq 0.
\eea
To check if this is the case, we count the longitudinal momentum fraction 
factors in Eq.~(\ref{jnon}), e.g. in the $q^+\to 0$ limit, the 1st term in
$I^{+nv}_{h'h}$ becomes 
\bea\label{zm}
I^{+{\rm z.m.}}_{h'h}&\sim&
\lim_{\al\to1}\int^\al_1\frac{dx}{xx''(x-\al)}
\frac{S^+_{h'h}(k^-=k^-_\Lambda)}
{(M^2_v-M^2_{0\Lambda f})(q^2-M^2_{\Lambda\Lambda})}
\nonumber\\
&=&\lim_{\al\to1}
\int^{\al}_{1}dx
\frac{(1-x)}{(1-\al)}[\cdots]S^{+}_{h'h}(k^-_\Lambda),
%\nonumber\\
%&=&\lim_{\al\to1}\int^{\al}_{1}dz (1-\al)(1-z)[\cdots]
%S^{+}_{h'h}(k^-_\Lambda),
\eea
where the factor $[\cdots]$ corresponds to the part that is
regular as $q^+\to 0$(i.e. $\al\to 1$). Thus, 
the nonvanishing zero-mode contribution is possible only if
the longitudinal momentum fraction of $S^+_{h'h}(k^-)$ 
is proportional to $(1-x)^{-s}$ with $s\geq 1$. Otherwise,
there is no zero-mode contribution 
since the integration range shrinks to zero as $q^+\to 0$. 
Other three terms in Eq.~(\ref{jnon}) have the same 
behavior as the first term shown in Eq.~(\ref{zm}).

Since we already know that the helicity zero-to-zero component,
$S^+_{00}$, gives a nonvanishing zero-mode contribution at the
level of simple vertex, $\Gamma^+=\gamma^+$~\cite{BCJ_rho}, 
we need to focus on other helicity components for the vector meson
vertex given by Eq.~(\ref{Gal}).
In the nonvalence region($P^+<k^+<P'^+$), since the LF energy poles,
$k^-_{m_1}$ and $k^-_{\Lambda_1}$, are proportional to $k^-\sim 1/(1-x)$,
we know that the $D$-terms of covariant and LFQM vertices in
Eq.~(\ref{Gal}) behave as
\be\label{D-term}
D_{\rm cov}\sim \frac{1}{1-x},
D_{\rm LFQM}\sim\sqrt{\frac{1}{1-x}},
\ee 
by counting only the singular longitudinal momenta in the 
$\al\to 1$(i.e. $x\to 1$) limit. We also note that the invariant
mass in the $D_{\rm LFQM}$ for the initial state vector meson 
has to be replaced by
\bea\label{M0f}
M^2_{0i}&=&\frac{m^2 + {\bf k}_{i\perp}}{x}
+ \frac{m^2 + {\bf k}_{i\perp}}{1-x}
\nonumber\\
&\to&
\frac{m^2 + {\bf k}_{i\perp}}{x}
+ \frac{m^2 + {\bf k}_{i\perp}}{x-1}
\eea
in the nonvalence region($x>1$) since the initial state vertex 
becomes the non-wavefunction vertex~\cite{JC01}(see Fig.~1(c)). 
Nevertheless, the power 
of the singular term in $1-x$ given in Eq.~(\ref{D-term}) remains
the same. 
Now, from the explicit forms of the trace terms $S^+_{ih'h}$
given in Appendix A, we could determine the existence/non-existence 
of the zero-mode for each helicity component.
Regardless of using covariant or LFQM vertices, we find that there are 
no singular terms in $1/(1-x)$ for the helicity ($++,+-,+0$) components.
The most problematic term regarding on the zero-mode was found to be 
the 2nd term $(S^+_{2+0})_{\rm off}$ given by Eq.~(\ref{Sp0_inst}), which
is proportional to $(k^--k^-_{\rm on})/D$.
While $(k^--k^-_{\rm on})/D_{\rm cov.}$ is regular in $1/(1-x)$, 
$(k^--k^-_{\rm on})/D_{\rm LFQM}\sim\sqrt{1/(1-x)}$, i.e. 
$S^+_{+0}$ itself shows singular behavior in
$\al\to 1$ limit for the LFQM vertex.
However, from Eq.~(\ref{zm}), the net result of the zero-mode contribution 
to $I^{+nv}_{+0}$ for $D=D_{\rm LFQM}$ is shown to be zero because  
\bea\label{s2p0}
I^{+{\rm z.m.}}_{+0}&\sim&
\lim_{\al\to1}\int^{1}_{\al}dx
\frac{(1-x)}{(1-\al)}[\cdots]\sqrt{\frac{1}{1-x}}
\nonumber\\
&=&\lim_{\al\to1}\int^{1}_{0}dz 
\frac{(1-\al)(1-z)}{\sqrt{(1-\al)(1-z)}}[\cdots]
\nonumber\\
&=& 0,
\eea
where the variable change $x=\al + (1-\al)z$ was made and the term 
$[\cdots]$ again corresponds to the regular part in $q^+\to 0$ limit. 
For helicity $(00)$ component, the zero-mode contribution comes from
$(S^+_{100})_{\rm off}$ for both $D=D_{\rm cov}$ and 
$D_{\rm LFQM}$(see Eq.~(\ref{S00_inst})). 

Thus, as in the case of Dirac coupling $\Gamma^\mu=\gamma^\mu$, 
we find that only $I^+_{00}$ receives the zero-mode contribution
for the more general vertex structures given by Eq.~(\ref{Gal})
with $D_{\rm cov.}$ and $D_{\rm LFQM}$.
Accordingly, we can compute the electromagnetic structure of the 
spin-1 particle using only the valence contribution 
as far as $I^+_{00}$ is avoided(e.g. GK prescription). 
Moreover, we may identify
the zero-mode contribution to $I^+_{00}$ using the angular condition
given by Eq.~(\ref{ac}), i.e.
\be\label{ac_Breit}
I^{+\rm z.m.}_{00}=
(1+2\kappa) I^{+\rm val}_{++} + I^{+\rm val}_{+-}
-\sqrt{8\kappa}I^{+\rm val}_{+0} - I^{+\rm val}_{00},
\ee
in the $q^+=0$ frame with the LF gauge $\ep^+_{\pm}=0$. 

Thus far, we relied on the generalization of the meson vertex(Eq.~(\ref{Gal}))
while we kept the use of smeared photon-vertex as adopted in 
Ref.~\cite{BCJ_rho}. Although the spin-orbit part of the LFQM can be 
incorporated by this generalization, the radial part of the LF wavefunction
given by Eq.~(\ref{chii}) may serve only semi-realistic calculation of 
physical observables as disscussed by others~\cite{FM,Ja03}.
For the more realistic calculation, one may need to replace the LF radial
wavefunction by the one which has much more phenomenological support.
For this purpose, we may utilize the LFQM presented in Ref.~\cite{CJMix},
which has been quite successful in predicting various static
properties of low lying hadrons.
Comparing $\chi$ in Eq.~(\ref{jjiv}) with our light-front
wave function given by Ref.~\cite{CJMix}, we get 
\be\label{LFvertex}
\chi_i(x,{\bf k}_{i\perp})=
\sqrt{\frac{8\pi^3}{N_c}}
\sqrt{\frac{\partial k_z}{\partial x}}
\frac{[x(1-x)]^{1/2}}{M_{0i}}\phi(x,{\bf k}_{i\perp}),
\ee
where the Jacobian of the variable tranformation
${\bf k}=(k_z,{\bf k}_\perp)\to
(x,{\bf k}_\perp)$ is obtained as $\partial k_z/\partial x=M_{0i}/[4x(1-x)]$
and the radial wave function is given by
\be\label{Radwf}
\phi({\bf k}^2)=\sqrt{\frac{1}{\pi^{3/2}\beta^3}}\exp(-{\bf k}^2/2\beta^2),
\ee
which is normalized as 
\be\label{norm}
\int d^3k|\phi({\bf k}^2)|^2
=\int^1_0 dx \int d^2{\bf k}_\perp
\biggl(\frac{\partial k_z}{\partial x}\biggr)
|\phi(x,{\bf k}_\perp)|^2 =1. 
\ee
In the next section, we shall investigate the $\rho$-meson
electromagnetic properties using this model.
We should note that this replacement of the radial part of the LF
wavefunction cannot alter our conclusion of zero-mode(i.e. only in 
$S^+_{00}$) because the use of Eq.~(\ref{LFvertex}) is limited just 
for the valence contribution.

In summary, the LFQM described above provides the calculation of physical
form factors with just the valence contribution of $I^+_{h'h}$ 
except the component of $(h',h)=(0,0)$; i.e.
\bea\label{I_LFQM}
I^{+{\rm LFQM}}_{h'h}&=&\int^1_0\frac{dx}{2(1-x)}\int d^2{\bf k}_\perp
\sqrt{\frac{\partial k_z}{\partial x}}
\sqrt{\frac{\partial k'_z}{\partial x}}
\nonumber\\
&\times& \phi(x,{\bf k}_{i\perp})\phi^*(x,{\bf k}_{f\perp})
\frac{(S^+_{h'h})_{\rm on}}{M_{0i}M_{0f}}.
\eea
Thus, in the GK prescription where $I^+_{00}$ is not used, it is
easy to verify $G_C^{GK}(0)=I^{+{\rm LFQM}}_{++}/2P^+=1$ because at $Q=0$,
$M^2_{0i}=M^2_{0f}=M^2_0=(m^2+{\bf k}^2_\perp)/x(1-x)$,
$(S^+_{++})_{\rm on}=4P^+(1-x)M^2_0$ from  a straightforward
evaluation using Eq.~(\ref{Spp_on})
and $(S^+_{+-})_{\rm on}=0$ from Eq.~(\ref{Spm_on}).

However, the BH prescription involves $I^+_{00}$ component and 
$G^{BH}_C(0)=I^{+{\rm LFQM}}_{00}/2P^+\neq 1$ due to the zero-mode contribution.
It has been shown in Ref.~\cite{KS}
that the additive model for the current operator of 
interacting constituents is consistent with the angular condition only
for the first two terms of the expansion of the plus(good) current in
powers of the momentum transfer $Q$.
Indeed, we can show that the zero-mode contribution to $I^{+{\rm LFQM}}_{00}$
vanishes in the zero-binding limit(i.e. $M_v=M_0$) and the angular
condition $\Delta(0)=0$ in this case. At the same token, we can show
that $I^{+{\rm LFQM}}_{++}=I^{+{\rm LFQM}}_{00}$(i.e. $(S^+_{++})_{\rm on}
=(S^+_{00})_{\rm on}$ from Eqs.~(\ref{Spp_on}) and~(\ref{S00_on}))
at $Q=0$ in the zero-binding limit.
In most previous LFQM analyses~\cite{Ja90,Card,DJ}, the authors
used not only the meson
vertex factor given by Eq.~(\ref{Gal}) together with Eq.~(\ref{aap_LF})
but also the zero binding energy prescription, i.e. $M_v=M_0$,
which is equivalent to replace $M_v$ in $\ep^\mu(P,0)[\ep^\mu(P',0)]$  with
$M_{0i}[M_{0f}]$.  In that case, the angular condition is zero at $Q^2=0$.
However, in this work, we showed an explicit example which gives 
$\Delta(0)\neq 0$ even if the plus current is used. 
In Ref.~\cite{BCJ_rho}, $\Delta(0)\neq 0$ was also shown.
Thus, the zero-mode 
contribution is in general necessary even at $Q^2=0$ for the $I^+_{00}$ 
amplitude.
\section{Numerical Results}
The model parameters used in our analysis are $m=0.22$ GeV and 
$\beta=0.3659$ GeV, which were obtained from 
the linear confining potential  of our QCD-motivated effective
Hamiltonian in LFQM~\cite{CJMix}, as well as $M_v=0.77$ GeV.

\begin{figure}
\includegraphics[height=7cm,width=7cm]{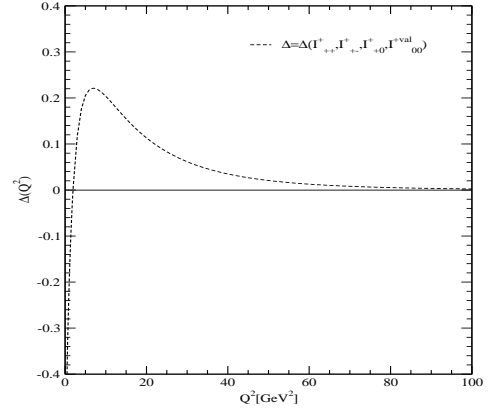}
\caption{ The angular condition given by Eq.~(\ref{ac}) or~(\ref{ac_Breit})
in the Breit frame, i.e.  $\Delta(Q^2)=I^{+\rm zm}_{00}$.
Note that the zero-mode contribution is necessary to satisfy the angular
condition even at $Q^2=0$, where $\Delta(0)=-0.65$ in our model calculation. 
\label{Fig_AC}}
\end{figure}
First, we show in Fig.~\ref{Fig_AC} the angular condition $\Delta(Q^2)$
given by Eq.~(\ref{ac}) or Eq.~(\ref{ac_Breit}) in the Breit frame 
defined by Eq.~(\ref{BF_F}).
In this particular reference frame, $\Delta(Q^2)$ is equal to the
zero-mode from the helicity zero-to-zero component, $I^{+\rm z.m.}_{00}$  
in Eq.~(\ref{ac_Breit}).
Note that $\Delta(Q^2)\neq 0$ even at $Q^2=0$(See the discussion
just before this section.). Unless the binding-energy
zero-limit($M_v=M_0$) is taken,  $I^+_{00}$ is not
immune to the zero-mode even at $Q^2=0$, but it eventually goes to zero
as $Q^2$ becomes very large. 

\begin{figure*}
\includegraphics[height=7cm,width=7cm]{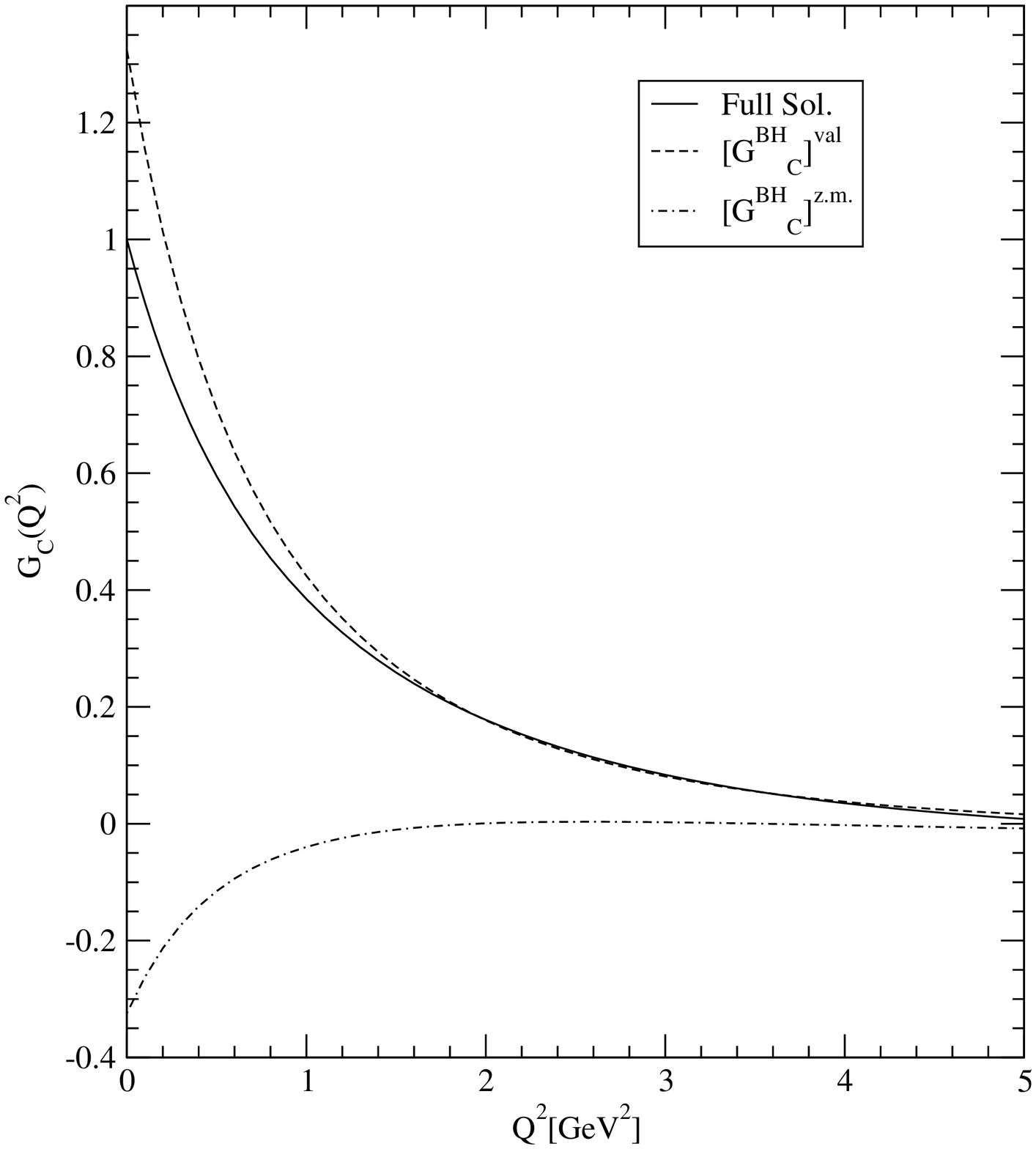}
\includegraphics[height=7cm,width=7cm]{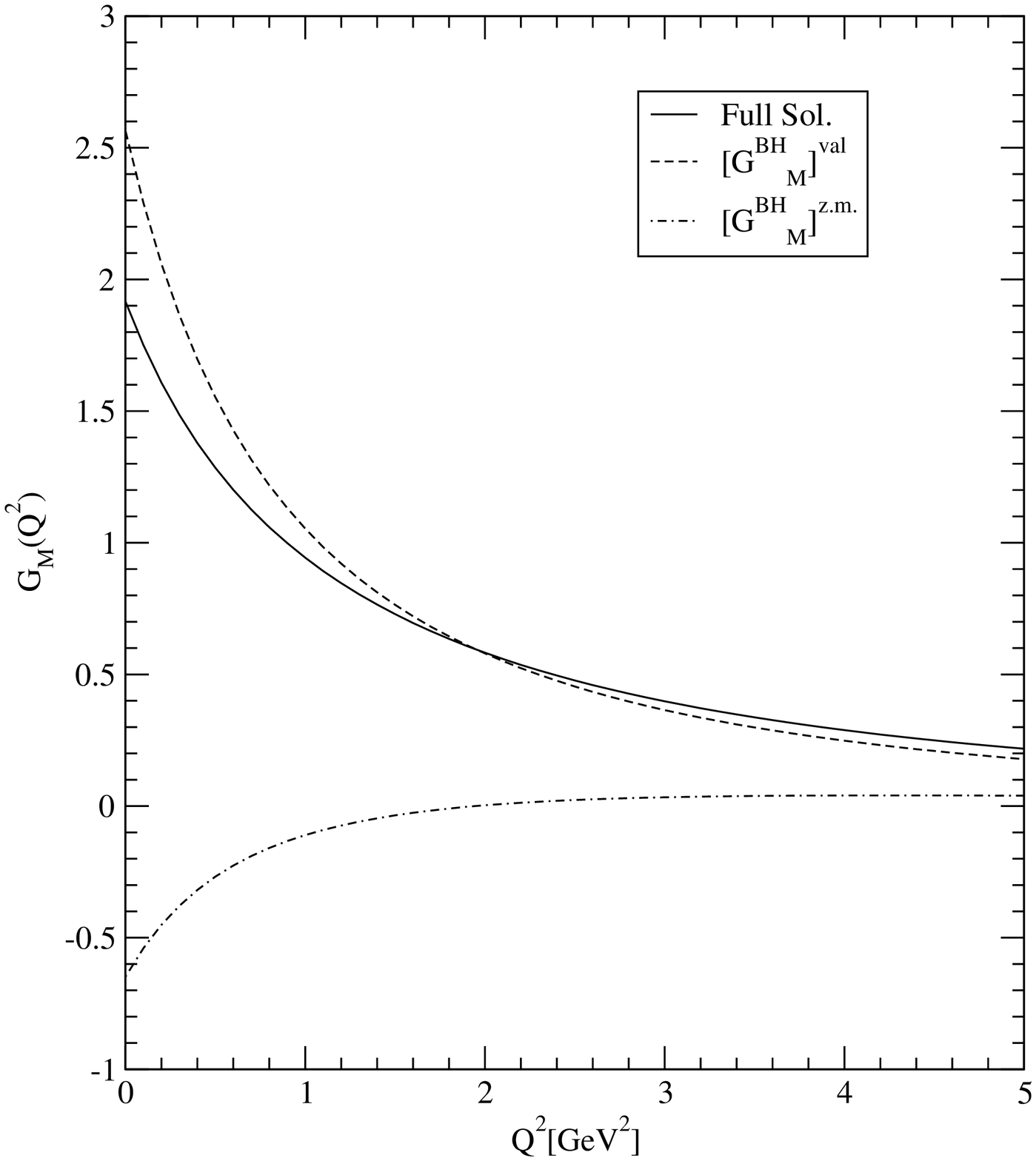}
\includegraphics[height=7cm,width=7cm]{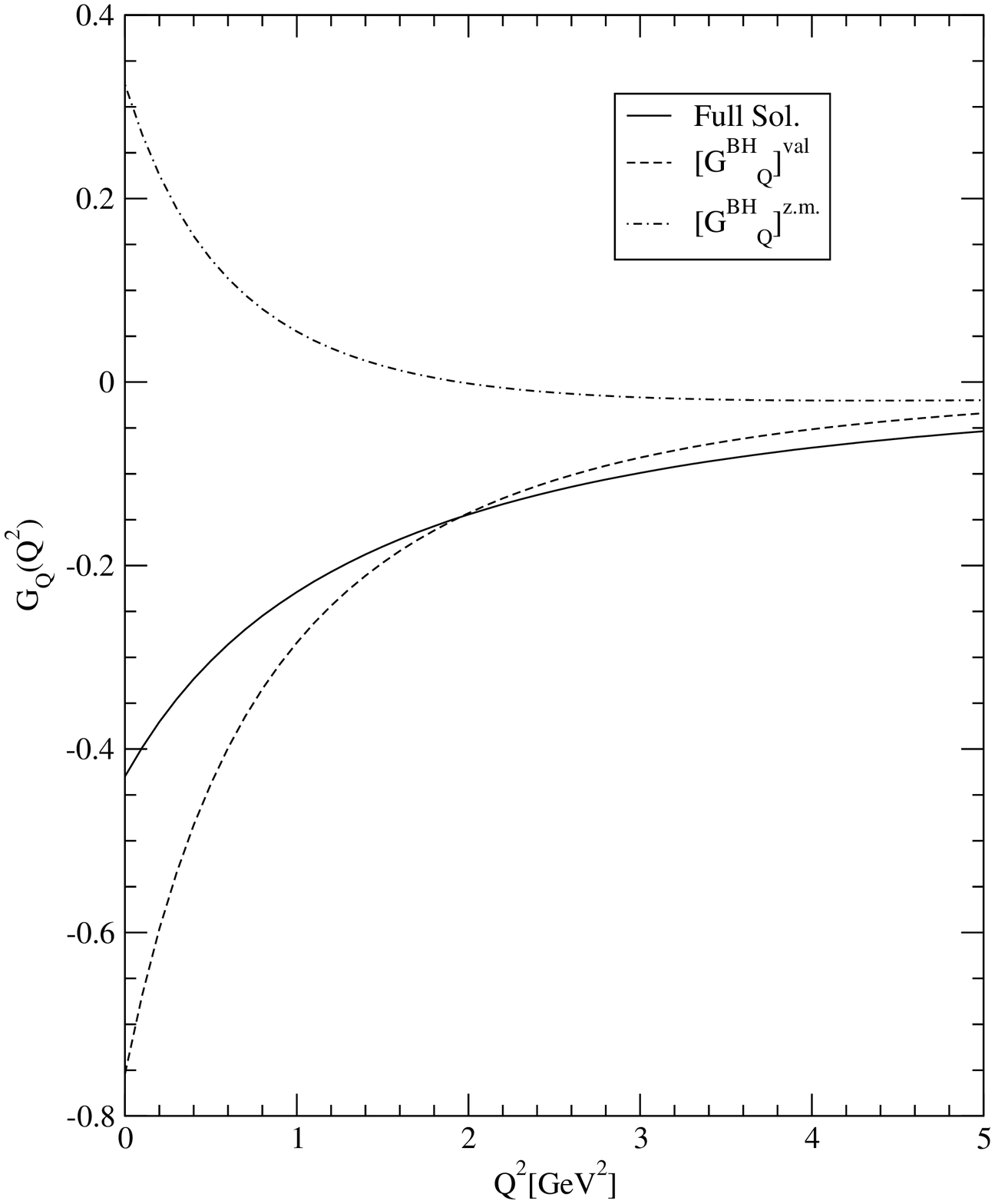}
\caption{Form factors of the $\rho$ meson, i.e. charge($G_C(Q^2)$),
magnetic($G_M(Q^2)$), and quadrupole($G_Q(Q^2)$) form factors.
The solid and dashed lines represent the full(i.e. valence+zero-mode)
solution and valence contribution($I^{+\rm val}_{00}$)
contribution to the form factors, respectively. 
The dash-dotted line represents the zero-mode contribution, i.e.
full solution $-$ valence contribution.\label{Fig_Form}}
\end{figure*}

In Fig.~\ref{Fig_Form}, we plot the physical form factors, $G_C$, $G_M$,
and $G_Q$, respectively. 
The solid lines represent the full solutions, i.e. GK
prescription in Eq.~(\ref{G_GK}) or equivalently BH prescription
in Eq.~(\ref{G_BH}) including the zero-mode contribution to $I^+_{00}$, i.e.
$I^{+{\rm full}}_{00}=I^{+{\rm val}}_{00}+I^{+{\rm z.m.}}_{00}$, where 
$I^{+{\rm z.m.}}_{00}$ is obtained from Eq.~(\ref{ac_Breit}).
As they should be, the full solutions are found to be independent from the
choice of prescription.
The dashed lines represent the valence contributions to the form factors
in the BH prescription.  The dashed-dotted lines represent 
the zero-mode from the helicity zero-to-zero component,
which turns out to be exactly the same with the difference between solid and
dashed lines.  
For the charge form factor, $G_C(Q^2)$, we found it has a zero around 
$Q^2=5.5$ GeV$^2$, which is not shown in the figure.
Also, the nonvanishing zero-mode contribution to $I^+_{00}$ 
at $Q^2=0$ is apparent in Fig.~\ref{Fig_Form}. 

We also obtain the magnetic(in unit of $e/2M_v$) and 
quadrupole(in unit of $e/M^2_v$) moments for the $\rho$ meson as
\begin{eqnarray*}
\mu=1.92,\; Q=0.43.
\end{eqnarray*}
Our result for the magnetic moment without involving the zero-mode
is comparable with $\mu=1.83$ in 
Ref.~\cite{Ja03} and the one in Ref.~\cite{SS}, in which the author found
$\mu<2$ by considering the low energy limit of the radiative amplitudes
in conjunction with the amplitude calculated by the hard-pion technique. 
The recent light-cone
QCD sum rule~\cite{Aliev} and the traditional QCD sum rule~\cite{AS} reported
$\mu=2.3\pm 0.5$ and $2.0\pm 0.3$, respectively. We note that the author of
Ref.~\cite{AS} also preffered $\mu<2$.
On the other hand, the previous LFQM~\cite{Card,DJ} and the 
Dyson-Schwinger model~\cite{DS1,DS2} predicted $\mu>2$. 
We summarize in Table~\ref{t1} our results of magnetic dipole($\mu$) and
quadrupole($Q$) moments compared with other theoretical predictions.

\begin{table}
\caption{Magnetic dipole($\mu$) and quadrupole($Q$) moments
in units of $e/2M_v$ and $e/M^2_v$, respectively. }
\begin{tabular}{|c|c|c|c|c|}
\hline
References & This work& ~\protect\cite{Ja03}
&~\protect\cite{Bag}&~\protect\cite{AS} \\
\hline
$\mu$ & 1.92 & 1.83 & 2.3 &2.0$\pm$0.3 \\
\hline
$Q$ & 0.43 & 0.33 & 0.45 & -\\
\hline
\end{tabular}
\label{t1}
\end{table}

\begin{figure*}[t]
\includegraphics[height=7cm,width=7cm]{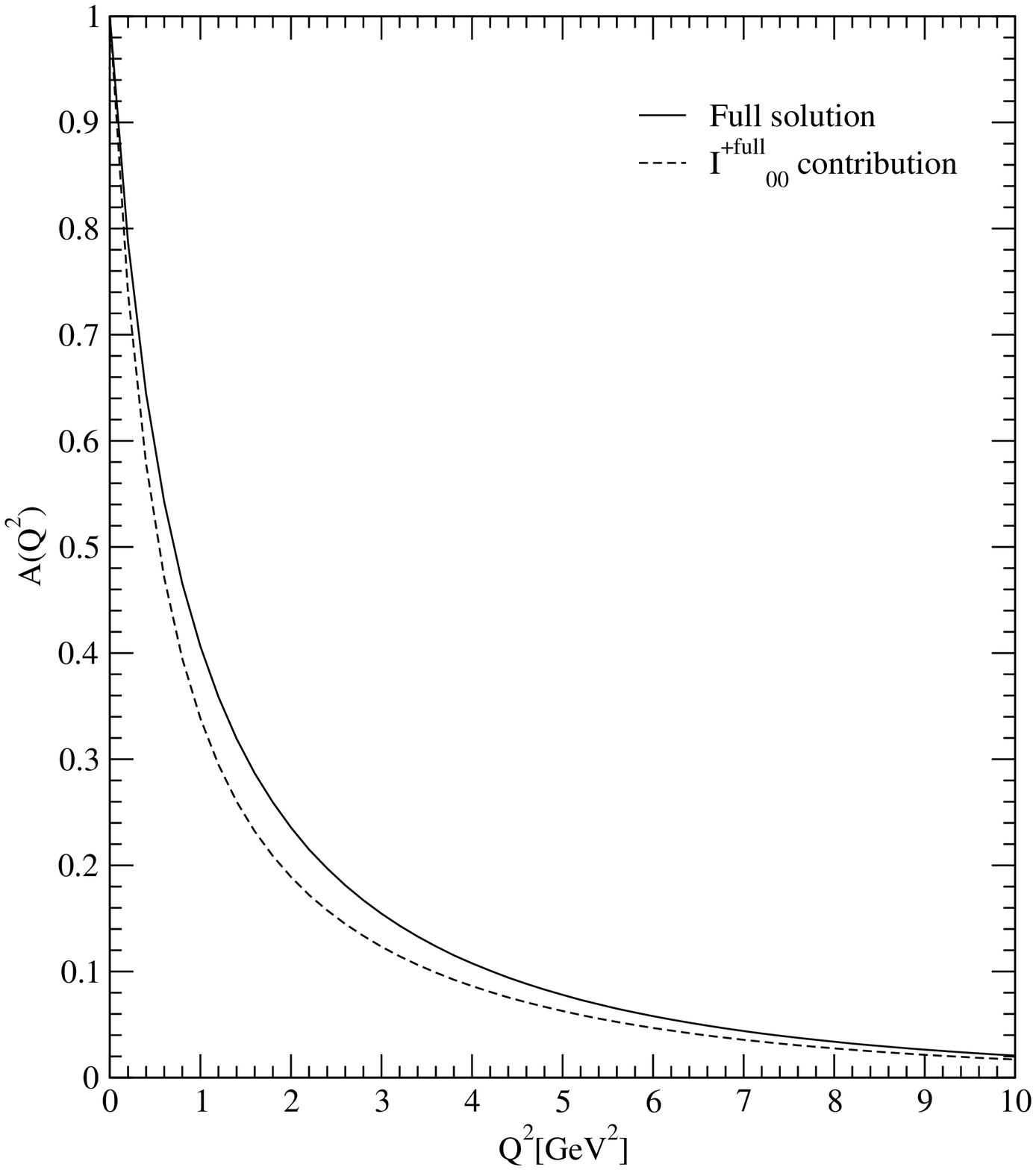}
\includegraphics[height=7cm,width=7cm]{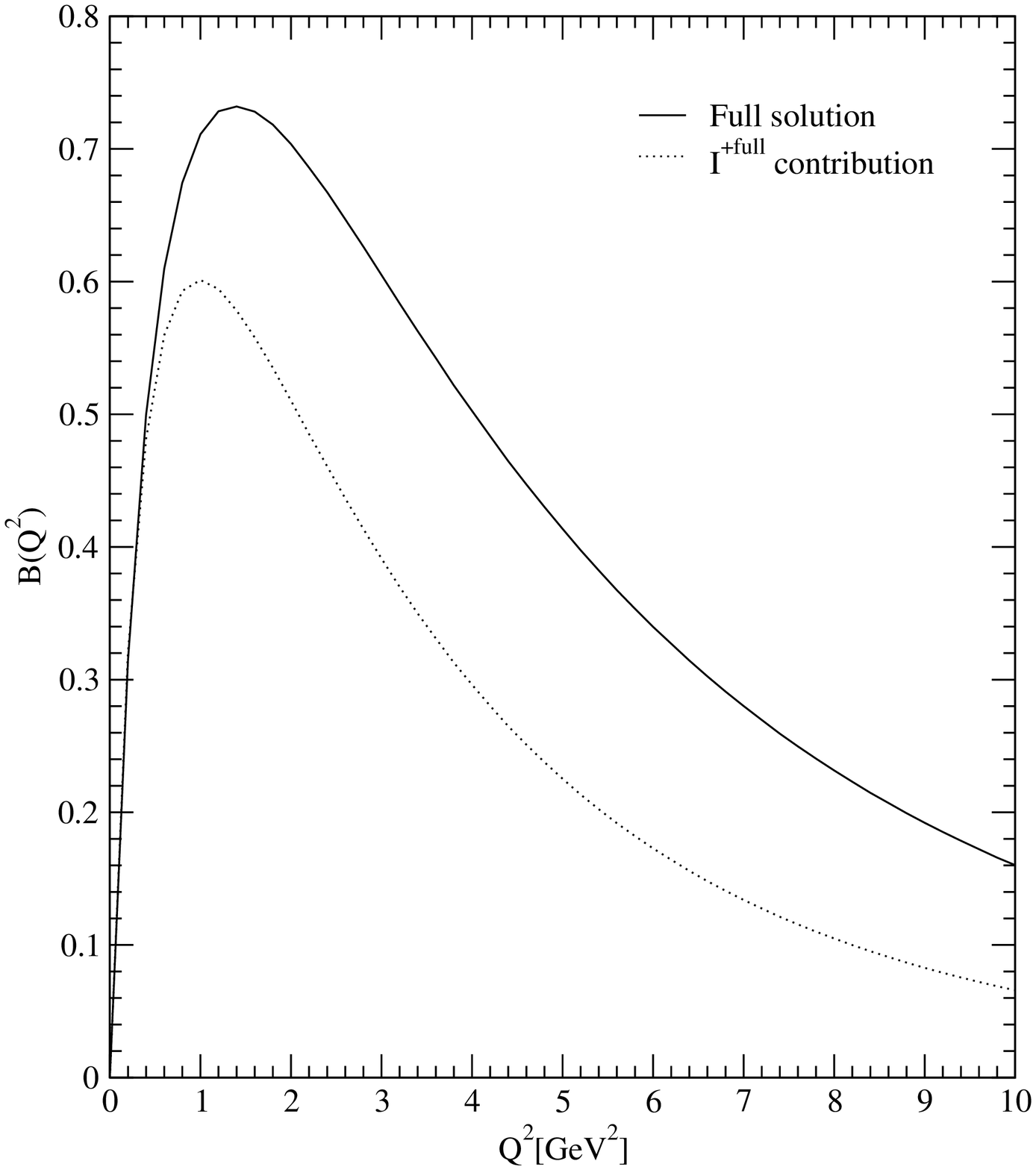}
\includegraphics[height=7cm,width=7cm]{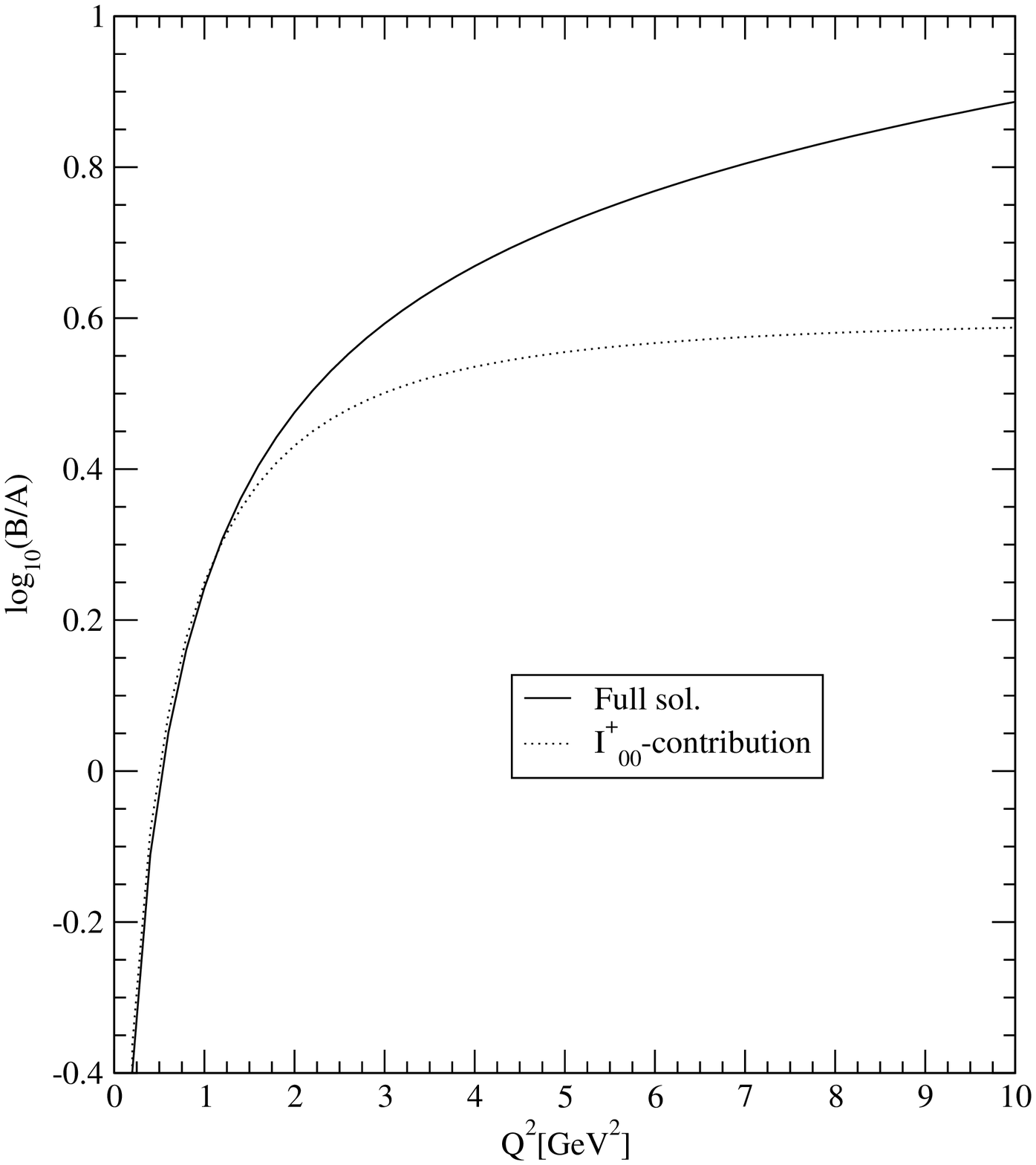}
\caption{Structure functions $A(Q^2)$, $B(Q^2)$, and their
ratio $\log_{10}(B/A)$. Solid and dotted lines represent the full results
and the contributions only from the 
$I^{+{\rm full}}_{00}=I^{+{\rm val}}+I^{+{\rm z.m.}}$
component.\label{Fig_AB}}
\end{figure*}
In Fig.~\ref{Fig_AB}, we plot the structure functions $A(Q^2)$, 
$B(Q^2)$ and their ratio $\log_{10}(B/A)$ up to $Q^2=10$ GeV$^2$. 
The solid and dashed lines represent the full results and the contributions
only from the $I^{+\rm full}_{00}(=I^{+{\rm val}}_{00}+I^{+{\rm z.m.}}_{00})$ 
component, respectively.  The dominance of helicity zero-to-zero amplitude
at high $Q^2$ region~\cite{BH,CaJi,CHH} has been discussed in the context of
PQCD counting rules and the naturalness condition~\cite{BJ1}.
However, the analysis of angular condition~\cite{CaJi} reveals that the
subleading contribution 
$I^+_{+0}\sim\frac{\Lambda_{QCD}}{Q}I^+_{00}$ and
$I^+_{++}\sim I^+_{+-}\sim\biggl(\frac{\Lambda_{QCD}}{Q}\biggr)^2 I^+_{00}$
are not as negligible as one might naively expect from PQCD.
Our LFQM results indicate that the dominance of $I^+_{00}$ is realized
in $A(Q^2)$ but not in $B(Q^2)$(or $\log(B/A)$), supporting the significance
of the subleading contribution discussed in Ref.~\cite{CaJi}.
For the magnetic form factor $G_M(Q^2)$ in Eq.~(\ref{G_BH}),
both $I^+_{00}$ and $I^+_{+0}$ terms contribute at the same order
and the $I^+_{+0}$ contribution at large $Q^2$ region(even at
$Q^2\sim 100$ GeV$^2$) is not negligible.
This explains why we have discrepancies between the full results and
$I^+_{00}$ dominances for the calculations of $B(Q^2)$ and 
$\log_{10}(B/A)$.

\begin{figure*}
\includegraphics[height=7cm,width=7cm]{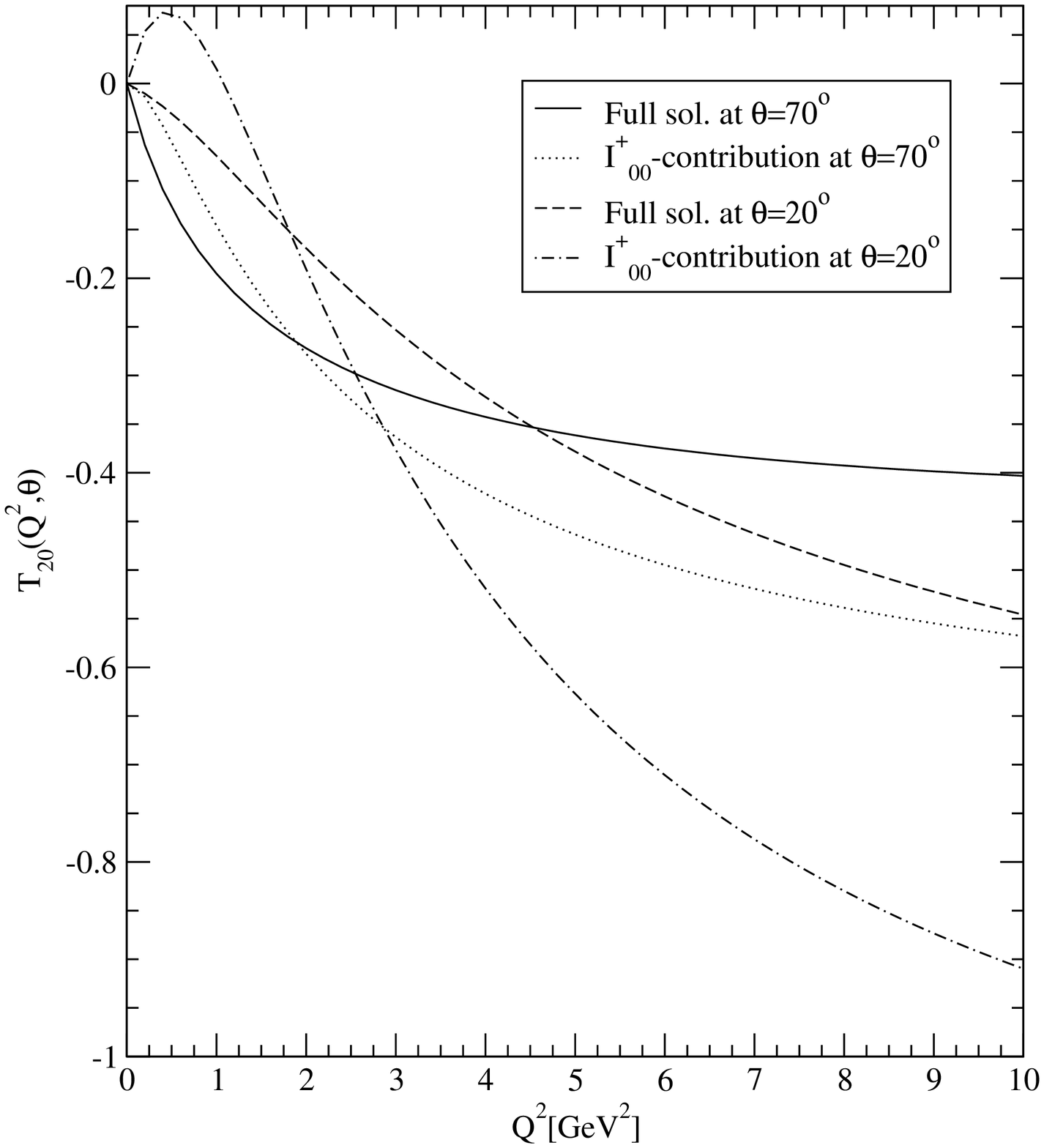}
\caption{Tensor polarization $T_{20}(Q^2)$ at $\theta=70^{\circ}$(solid 
line) and $\theta=20^{\circ}$(dashed line). For comparison, we also
show the $I^+_{00}$ dominance at $\theta=70^{\circ}$(dotted line)
and $\theta=20^{\circ}$(dot-dashed line). \label{Fig_T20}}
\end{figure*}

In  Fig.~\ref{Fig_T20}, we show the full solution of the
tensor polarization $T_{20}(Q^2)$ at $\theta=20^{\circ}$(dashed line) 
and at $\theta=70^{\circ}$(solid line), respectively.
For comparison, we also plot the $I^{+\rm full}_{00}$ contribution
to $T_{20}(Q^2)$ at $\theta=20^{\circ}$(dot-dashed line)
and at $\theta=70^{\circ}$(dotted line), respectively.
The $T_{20}(Q^2)$ results also indicate the non-negligible subleading 
contribution~\cite{CaJi} even at a rather large $Q^2$ region.

\section{Conclusion}
In this work, we investigated the electromagnetic structure of the $\rho$
meson in the light-front quark model. The two vector meson vertices
are analyzed, i.e. the manifestly covariant 
vertex $D=D_{\rm cov.}$ and the LFQM vertex $D=D_{\rm LFQM}$ consistent
with the Melosh transformation.
Using the power counting method for each helicity amplitude, we found that
only the helicity zero-to-zero amplitude receives the zero-mode contribution
regardless of $D=D_{\rm cov}$ or $D=D_{\rm LFQM}$.
Other helicity amplitudes such as $(h',h)=(+,+),(+,-)$ and $(+,0)$ are found
to be immune to the zero-mode for both vertices. 
Our finding is different from that in Ref.~\cite{Ja03},
in which the author found the nonvanishing zero-mode contribution to
the helicity zero-to-plus component as well~$^1$.
This is significant because the absence of zero-mode contribution in
$I^+_{+0}$ allows that the pure valence contribution in LFD can give
the full result of the physical form factors.

Further, we identified the zero-mode contribution to $I^+_{00}$ using the
angular condition given by Eq.~(\ref{ac_Breit}). Our result of
$\Delta(Q^2)$ exhibits the nonvanishing zero-mode even at $Q^2=0$ unless
the binding-energy zero limit($M_v=M_0$) is taken. This does not 
contradict the findings in the additive model for the current operator of
interacting constituents discussed in Ref.~\cite{KS}.
Indeed, our calculation with $D=D_{\rm LFQM}$ for $(h',h)=(+,+)$ and $(+,-)$
are equivalent to the previous LFQM~\cite{Card,DJ} calculations based on
the Melosh transformation. However, our calculations involving helicity
zero polarization vector, e.g. $(h',h)=(+,0)$, cannot be the same with
the previous Melosh-based calculations~\cite{Card,DJ} unless the zero
binding limit $M_v=M_0$ is taken.
Thus, it appears important to analyze the difference in the physical
observables including the binding energy effects.
As we presented in this work, our phenomenological  LFQM prediction 
including the binding
energy effect(i.e. $M_v\neq M_0$) leads to $\mu=1.92<2$, which is rather
different from the previous results($\mu>2$) based on the free Melosh
transformation~\cite{Card,DJ,Bag}.

Finally, our numerical results on $B(Q^2)$ and $T_{20}(Q^2)$ at large
$Q^2$ region support the significance of the subleading contribution,
 e.g. $I^+_{+0}$ contribution in $B(Q^2)$, found from the analysis
of the angular condition in Ref.~\cite{CaJi}.

\acknowledgements
We thank Ben Bakker for many helpful discussions. This work was supported 
in part by the grant from the U.S. Department of Energy 
(DE-FG02-96ER40947), the National Science Foundation (INT-9906384) and the 
Korea Research Foundation(KRF-2003-003-C00038). 
H.M.C would like to thank the staff of the Department of Physics at NCSU 
for their kind hospitality.
The National Energy Research Scientific 
Computer Center is also acknowledged for the grant of computing time.

\appendix
\begin{widetext}
\section{Summary of Trace terms in helicity amplitudes}
In the Breit frame with the LF gauge given by 
Eqs.~(\ref{BF_F}) and~(\ref{BD}), the explicit forms of the
trace terms in Eqs.~(\ref{Sji_on}) and~(\ref{Sji_inst}) for each
helicity component are given as follows.

(I) helicity ($++$)-component:
\bea\label{Spp_on}
( S^+_{1++})_{\rm on}&=&\frac{4P^+}{x}
\biggl[ m^2 + (2x^2-2x+1)\biggl({\bf k}^2_\perp - \frac{x^2}{4}Q^2
-ixk_yQ\biggr)\biggr],
\nonumber\\
( S^+_{2++})_{\rm on}&=&-2P^+\biggl(\frac{m}{D}\biggr)
\biggl[ 8(1-x){\bf k}^2_\perp + x(2x^2-2x+1)Q^2 + 2k_xQ 
         + 2ik_yQ(2x-1)^2 \biggr],
\nonumber\\
( S^+_{3++})_{\rm on}&=&-2P^+\biggl(\frac{m}{D'}\biggr)
\biggl[ 8(1-x){\bf k}^2_\perp + x(2x^2-2x+1)Q^2 - 2k_xQ 
        + 2ik_yQ(2x-1)^2 \biggr],
\nonumber\\
( S^+_{4++})_{\rm on}&=&\frac{4P^+}{DD'}
\biggl[ {\bf k}^2_\perp - \frac{x^2}{4}Q^2 - ixk_yQ\biggr]
\biggl[ (1-x)(M^2_{0i}+M^2_{0f}-8m^2)-xQ^2 \biggr]
\eea
and 
\bea\label{Spp_inst}
(S^+_{1++})_{\rm off}&=& 4(P^+)^2(k^- - k^-_{\rm on})(1-x)^2,
\nonumber\\
(S^+_{2++})_{\rm off}
&=&(S^+_{3++})_{\rm off}=0,
\nonumber\\
(S^+_{4++})_{\rm off}&=&
\frac{8(P^+)^2}{DD'}(1-x)^2(k^--k^-_{\rm on})
\biggl[ {\bf k}^2_\perp -\frac{x^2}{4}Q^2-ixk_yQ\biggr].
\eea
In the nonvalence region where $k^-\sim 1/(1-x)$, 
$D=D_{\rm cov.}\sim 1/(1-x)$ and $D=D_{\rm LFQM}\sim\sqrt{1/(1-x)}$. Thus,
we find from the power counting for the longitudinal momentum fraction 
that all the off-shell trace terms $(S^+_{i++})_{\rm off}(i=1,2,3,4)$ are
regular as $q^+\to 0$(or equivalently $x\to 1$). Therefore, there is no
zero-mode in the helicity ($++$) component.

(II) helicity ($+-$)-component:
\bea\label{Spm_on}
(S^+_{1+-})_{\rm on}&=& 2P^+(1-x)[ 4(k^L)^2-x^2Q^2],
\nonumber\\
(S^+_{2+-})_{\rm on}&=& 2P^+\biggl(\frac{m}{D}\biggr)
(2k^L+xQ)[ (2x^2-2x+1)Q + 4(1-x)k^L],
\nonumber\\
( S^+_{3+-})_{\rm on}&=& -2P^+\biggl(\frac{m}{D'}\biggr)
(2k^L-xQ)[ (2x^2-2x+1)Q -4(1-x)k^L],
\nonumber\\
( S^+_{4+-})_{\rm on}&=&-\frac{P^+}{DD'}
[4(k^L)^2 - x^2Q^2][(1-x)(M^2_{0i}+M^2_{0f}-8m^2)-xQ^2]
\eea
and
\bea\label{Spm_inst}
(S^+_{1+-})_{\rm off}&=&(S^+_{2+-})_{\rm off}
=(S^+_{3+-})_{\rm off} = 0,
\nonumber\\
(S^+_{4+-})_{\rm off}&=&-\frac{2(P^+)^2}{DD'}(k^- - k^-_{\rm on})
(1-x)^2[ 4(k^L)^2 -x^2Q^2].
\eea

Again, from the power counting for the longitudinal momentum fraction, 
$(S^+_{4+-})_{\rm off}$ is regular as $x\to 1$. Therefore, there is no
zero-mode in the helicity $(+-)$ component. 

(III) helicity ($+0$)-component:
\bea\label{Sp0_on}
(S^+_{1+0})_{\rm on}&=& \frac{P^+\sqrt{2}}{M_v}
(2k^L-xQ)(2x-1)(1-x)(M^2_{0i}+M^2_v),
\nonumber\\
(S^+_{2+0})_{\rm on}&=&-\frac{4P^+}{M_v\sqrt{2}}
\biggl(\frac{m}{D}\biggr)[(1-x)M^2_{0i}-xM^2_v]
[(2x^2-2x+1)Q+ 4(1-x)k^L],
\nonumber\\
( S^+_{3+0})_{\rm on}&=& \frac{4P^+}{M_v\sqrt{2}}
\biggl(\frac{m}{D'}\biggr)(2x-1)(1-x)(M^2_{0i}+M^2_v)(2k^L -xQ),
\nonumber\\
( S^+_{4+0})_{\rm on}&=&\frac{P^+\sqrt{2}}{M_v DD'}
(2k^L-xQ)[(1-x)M^2_{0i}-xM^2_v]
[(1-x)(M^2_{0i}+M^2_{0f}-8m^2)-xQ^2],
\eea
and 
\bea\label{Sp0_inst}
(S^+_{1+0})_{\rm off}&=& -\frac{4(P^+)^2}{M_v\sqrt{2}}
(k^- -k^-_{\rm on})(1-x)(2k^L + xQ),
\nonumber\\
(S^+_{2+0})_{\rm off}&=& 
-\frac{4(P^+)^2}{M_v\sqrt{2}}\biggl(\frac{m}{D}\biggr)
(k^- - k^-_{\rm on})[ (2x^2-2x+1)Q + 4(1-x)k^L],
\nonumber\\
(S^+_{3+0})_{\rm off}&=& -\frac{4(P^+)^2}{M_v\sqrt{2}}
\biggl(\frac{m}{D'}\biggr)(k^- -k^-_{\rm on})(1-x)(2k^L-xQ),
\nonumber\\
(S^+_{4+0})_{\rm off}&=& 
\frac{(P^+)^2\sqrt{2}}{M_v DD'}(k^- - k^-_{\rm on})(2k^L-xQ)
[(1-x)(M^2_{0i}+M^2_{0f}-8m^2)-xQ^2]
\nonumber\\
&&+\frac{2\sqrt{2}(P^+)^2}{M_v DD'}
(k^- -k^-_{\rm on})(1-x)^2(2k^L - xQ)[ (1-x)M^2_{0i}-xM^2_v]
\nonumber\\
&&+\frac{2\sqrt{2}(P^+)^3}{M_v DD'}(k^- -k^-_{\rm on})^2(1-x)^2
(2k^L -xQ).
\eea
From the power counting of the longitudinal momentum fraction in the
nonvalence region, we find only $(S^+_{2+0})_{\rm off}$ for $D=D_{\rm LFQM}$
shows singular behavior, i.e 
$(S^+_{2+0})_{\rm off}\sim\sqrt{1/(1-x)}$ as $x\to 1$.
However, as we show in Eq.~(\ref{s2p0}), 
the resulting zero-mode contribution vanishes due to the prefactor 
involved in the integration. Therefore, there is no zero-mode contribution
to the helicity $(+0)$ component.

(IV) helicity (00)-component:
\bea\label{S00_on}
( S^+_{100})_{\rm on}&=&\frac{4P^+}{M^2_v}x(1-x)^2
(M^2_{0i}+M^2_v)(M^2_{0f}+M^2_v),
\nonumber\\
(S^+_{200})_{\rm on}&=& \frac{4P^+}{M^2_v}
\biggl(\frac{m}{D}\biggr)(2x-1)(1-x)(M^2_{0f}+M^2_v)
[(1-x)M^2_{0i}-xM^2_v],
\nonumber\\
(S^+_{300})_{\rm on}&=& \frac{4P^+}{M^2_v}
\biggl(\frac{m}{D'}\biggr)(2x-1)(1-x)(M^2_{0i}+M^2_v)
[(1-x)M^2_{0f}-xM^2_v],
\nonumber\\
(S^+_{400})_{\rm on}&=& \frac{2P^+}{M^2_v DD'}
[(1-x)M^2_{0i}-xM^2_v][(1-x)M^2_{0f}-xM^2_v]
[(1-x)(M^2_{0i}+M^2_{0f}-8m^2)-xQ^2],
\eea
and
\bea\label{S00_inst}
(S^+_{100})_{\rm off}&=&\frac{4(P^+)^2}{M^2_v}
(k^- -k^-_{\rm on})[m^2 + {\bf k}^2_{\perp}-x^2Q^2/4],
\nonumber\\
(S^+_{200})_{\rm off}&=&
\frac{4(P^+)^2}{M_v^2}\biggl(\frac{m}{D}\biggr)
(k^- - k^-_{\rm on})(1-x)(2x-1)(M^2_{0f}+M^2_v)
\nonumber\\
&&-\frac{4(P^+)^2}{M^2_v}
\biggl(\frac{m}{D}\biggr)(k^- -k^-_{\rm on})(1-x)
[(1-x)M^2_{0i}-xM^2_v + (k^- - k^-_{\rm on})P^+],
\nonumber\\
(S^+_{300})_{\rm off}&=&
\frac{4(P^+)^2}{M_v^2}\biggl(\frac{m}{D'}\biggr)
(k^- - k^-_{\rm on})(1-x)(2x-1)(M^2_{0i}+M^2_v)
\nonumber\\
&&-\frac{4(P^+)^2}{M^2_v}
\biggl(\frac{m}{D'}\biggr)(k^- -k^-_{\rm on})(1-x)
[(1-x)M^2_{0f}-xM^2_v + (k^- - k^-_{\rm on})P^+],
\nonumber\\
(S^+_{400})_{\rm off}&=&
\frac{2(P^+)^2}{M^2_v DD'}(k^- -k^-_{\rm on})
[(1-x)(M^2_{0i}+M^2_{0f})- 2xM^2_v]
[(1-x)(M^2_{0i}+M^2_{0f}-8m^2)-xQ^2]
\nonumber\\
&&+\frac{2(P^+)^3}{M^2_v DD'}(k^- -k^-_{\rm on})^2
[(1-x)(M^2_{0i}+M^2_{0f}-8m^2)-xQ^2]
\nonumber\\
&&+\frac{4(P^+)^2}{M^2_vDD'}(k^- -k^-_{\rm on})
(1-x)^2\biggl\{[(1-x)M^2_{0i}-xM^2_v][(1-x)M^2_{0f}-xM^2_v] 
\nonumber\\
&&\hspace{4cm}
+ P^+(k^- - k^-_{\rm on})[(1-x)(M^2_{0i}+M^2_{0f})-2xM^2_v]
+ (P^+)^2(k^- -k^-_{\rm on})^2\biggr\}.
\eea
From the power counting of the longitudinal momentum fraction, only
$(S^+_{100})_{\rm off}$ is singular for both $D_{\rm cov.}$ and
$D_{\rm LFQM}$, i.e. $(S^+_{100})_{\rm off}\sim 1/(1-x)$ as $x\to 1$, which
gives the zero-mode contribution to the helicity zero-to-zero component.
\end{widetext}

\end{document}